# NEXT-GENERATION INFORMATION TECHNOLOGY SYSTEMS FOR FAST DETECTORS IN ELECTRON MICROSCOPY

Dieter Weber, Alexander Clausen, Rafal E. Dunin-Borkowski

*Ernst Ruska-Centre for Microscopy and Spectroscopy with Electrons, Forschungszentrum Jülich*

*2019-03-20*

## Introduction

The Gatan K2 IS direct electron detector (Gatan Inc., 2018), which was introduced in 2014, marked a watershed moment in the development of cameras for transmission electron microscopy (TEM) (Pan & Czarnik, 2016). Its pixel frequency, i.e. the number of data points (pixels) recorded per second, was two orders of magnitude higher than the fastest cameras available only five years before. Starting from 2009, the data rate of TEM cameras has outpaced the development of network, mass storage and memory bandwidth by almost two orders of magnitude (Figure 1, Figure 2). Consequently, solutions based on personal computers (PCs) that were adequate until then are no longer able to handle the resulting data rates. Instead, tailored high-performance setups are necessary. Similar developments have occurred for advanced X-ray sources such as the European XFEL, requiring special information technology (IT) systems for data handling (Sauter, Hattne, Grosse-Kunstleve, & Echols, 2013) (Fangohr, et al., 2018). Information and detector technology are currently under rapid development and involve disruptive technological innovations. This chapter briefly reviews the technological developments of the past 20 years, presents a snapshot of the current situation at the beginning of 2019 with many practical considerations, and looks forward to future developments.

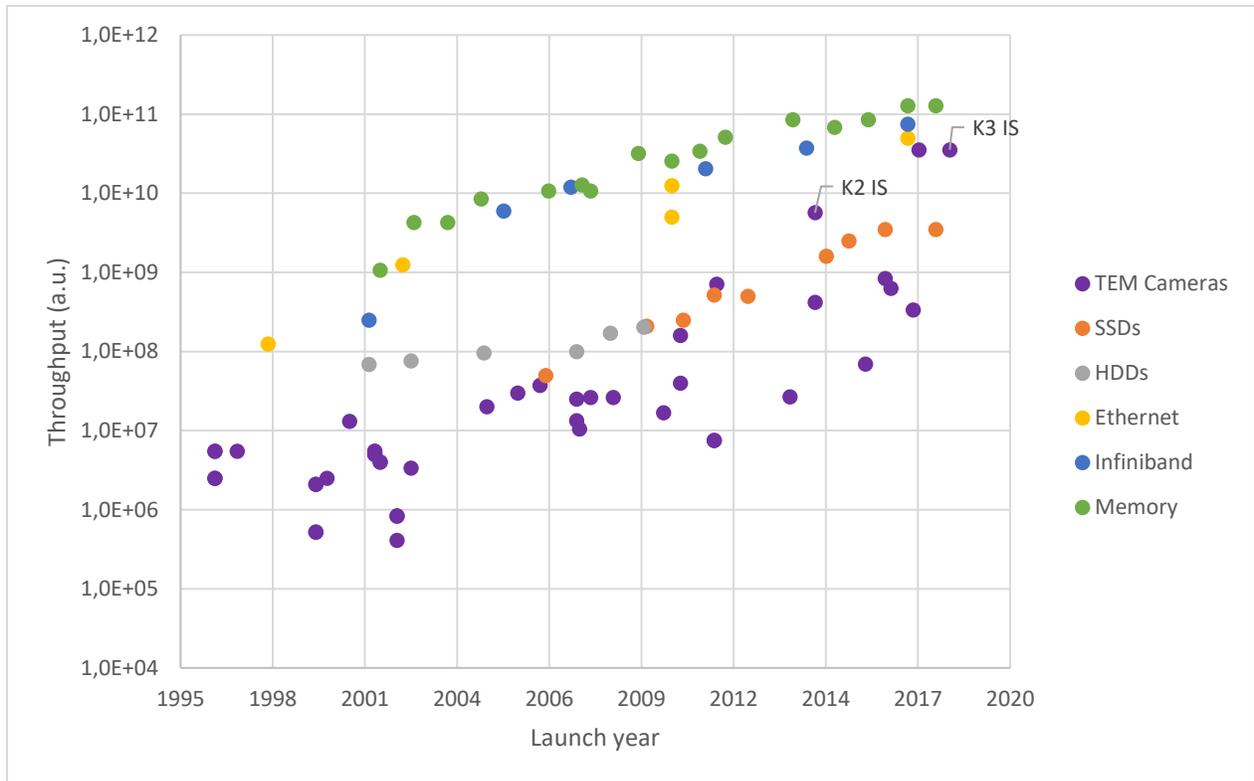

**FIGURE 1: EVOLUTION OF TEM CAMERA, SOLID-STATE DISK (SSD), HARD DISK DRIVE (HDD), ETHERNET, INFINIBAND AND MEMORY THROUGHPUT OVER TIME. THE THROUGHPUT IS GIVEN IN BYTES PER SECOND WHERE THIS INFORMATION IS AVAILABLE. FOR CAMERAS, THE PIXEL FREQUENCY IS GIVEN IN PIXELS PER SECOND. THE SOURCE FILE (WEBER, 2018) IS AVAILABLE ONLINE AND DESCRIBES IN DETAIL HOW THE DATA WAS COLLECTED AND EVALUATED.**

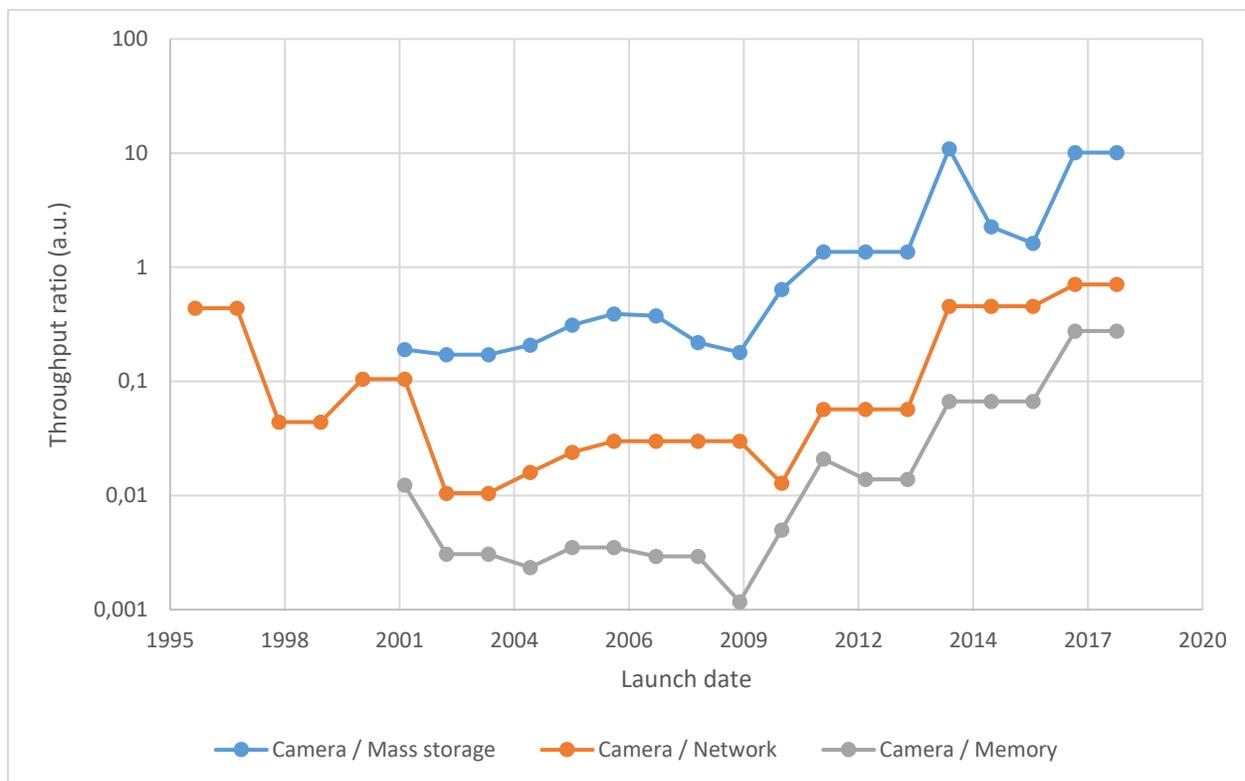

**FIGURE 2: DEVELOPMENT OF TEM CAMERA PIXEL FREQUENCY RELATIVE TO TYPICAL MASS STORAGE, NETWORK AND MEMORY THROUGHPUT BASED ON THE DATA IN FIGURE 1 (WEBER, 2018). THE RELATIVE DEVELOPMENT OF CAMERA PIXEL FREQUENCY IN COMPARISON TO THE DATA RATE OF MASS STORAGE (SSDS AND HDDS), ETHERNET AND MEMORY IS HIGHLIGHTED BY CALCULATING THEIR QUOTIENT. THE PLOT SHOWS THAT CAMERAS HAVE INCREASED THEIR SPEED BY ABOUT TWO ORDERS OF MAGNITUDE RELATIVE TO OTHER IT COMPONENTS SINCE 2009.**

## Application of high-speed detectors in TEM

In TEM, the first direct electron detectors for recording images were developed primarily to collect data for cryo electron microscopy (cryo EM), in order to increase the detection quantum efficiency by collecting data from individual electrons rather than integrating the dose of many electrons per pixel (Li, et al., 2013). Fast detectors with somewhat lower number of pixels and very high frame rates were developed for pixelated scanning TEM (pixelated STEM) (Simson, et al., 2015) (Yang, et al., 2015) (Tate, et al., 2016), where a full electron diffraction pattern is recorded for each position of a STEM scan. It has long been known that collections of electron diffraction patterns, in particular from overlapping sample regions, contain a wealth of information about the sample (Cowley, 1979) (Hegerl & Hoppe, 1970) (Hoppe, 1969) (Humphreys, Eaglesham, Maher, & Fraser, 1988) (Nellist, McCallum, & Rodenburg, 1995) (Steeds, 1979). Such fast detectors provide a practical route to tap into this information for real-world applications (Jiang, et al., 2018) (Krajnak, McGrouther, Maneuski, Shea, & McVitie, 2016) (MacLaren, et al., 2016) (Yang, Pennycook, & Nellist, 2015) (Nguyen, et al., 2016) (Ophus, Ercius, Sarahan, Czarnik, & Ciston, 2014) (Pennycook, et al., 2015) (Sagawa, et al., 2017), because even a small scan of 256 x 256 pixels would take almost two hours to record at 10 frames per second (fps), a speed that was considered state-of-the-art for high-speed TEM cameras until 2010 (Weber, 2018). By using the 400 fps of a Gatan K2 IS, the scan time is reduced to less than three minutes, while the 2400 fps of a Quantum Detectors Merlin camera (Quantum Detectors, 2017) reduces this time to below 30 s. Other applications of such high-speed cameras include the observation of dynamic processes during in situ experiments (Pan & Czarnik, 2016) (Lawrence, Chang, & Crozier, 2017).

## Challenges and requirements

Irrespective of their application, using fast electron detectors at their full rates generates large amounts of raw data. This situation creates significant challenges for data handling and processing (Belianinov, et al., 2015) (Jesse, et al., 2016). One option is to immediately reduce the data in a suitable fashion and to handle only a reduced data stream. This approach is often followed in cryo EM, where frames with individual electron read-outs are combined to some degree. However, users of pixelated STEM and high-speed in situ observations benefit immensely from keeping all raw data, so that it can be re-analyzed using different algorithms or settings as desired (Jiang, et al., 2018) (Krajnak, McGrouther, Maneuski, Shea, & McVitie, 2016) (MacLaren, et al., 2016) (Yang, Pennycook, & Nellist, 2015) (McCallum & Rodenburg, 1993). A data set from a single pixelated STEM scan can already potentially reach a size of up to 64 TB when using a high-resolution (1k x 1k) scan with high detector resolution (4k x 4k) recording float32 values. Practical pixelated STEM scans acquired with the Gatan K2 IS detector at the Ernst Ruska Centre in Jülich already reach sizes of 200 GB, and a session with several scans can produce 2.5 TB of data. These amounts are expected to grow rapidly as faster cameras, storage and processing systems become available.

In order to make full use of these capabilities, processing and storage have to move to dedicated systems because the data volumes and calculations are beyond what is convenient to handle on individual PCs and hard drives. A first glance at 2.5 TB of data stored on a typical fast PC storage medium using a 300 MB/s read rate would take more than two hours. A time frame of minutes or seconds instead of hours would be desirable for fast turn-around from an experiment to analysis, requiring read rates in the range of 100 GB/s to TB/s.

Furthermore, the acquisition systems themselves have to handle high data rates. For the biggest and fastest detectors, the readout for one detector must be distributed over several PCs, for example by dividing the detector into tiles with parallel readout (X-Spectrum). Such a high aggregate throughput creates challenges in data transfer from the readout location to a permanent storage, typically requiring a data center close to the acquisition site. Usual PC-based control systems cannot handle such a data rate. Acquisition systems should therefore provide a strongly reduced monitoring signal to the acquisition control interface as a feedback to the user. In view of these requirements, a suitable acquisition system is a complex distributed setup.

Electron microscopists are used to obtaining immediate visual feedback from the instrument. For pixelated STEM it is no longer the microscope's electron-optical system or simple electronics that create the image directly on a phosphor screen or monitor, but digital processing of the detector data (Li, Dyck, Kalinin, & Jesse, 2018). Real-time turnaround with processing of results is required to ensure fast analysis and optimal use of instrument time. Since the throughput of a typical computer is fundamentally limited by its memory bandwidth – currently between 40 GB/s (Intel, 2018) and 160 GB/s (Wikichip.org, 2018) – processing has to be distributed across between 10 and 1000 nodes even when using the fastest algorithms to achieve the TB/s data rates that are required for immediate feedback (Nielsen, 1994).

Given such requirements, performance and scalability are becoming key enablers for dealing with data. Speedup results in cost savings or a significant increase in capabilities, for example by being able to use larger data volumes or more complex computations (Delvecchio, 2011). In contrast to many previous analysis tasks, for which suitable response times could be achieved without special effort, this situation makes investments in performance optimization worthwhile (Busch, 2016).

Many previous supercomputing applications have involved running complex iterative simulations, such as weather models (Bauer, Thorpe, & Brunet, 2015), fluid dynamics calculations (Tezduyar, et al., 1996) and molecular models (Sanbonmatsu & Tung, 2007) very quickly. The emerging data analysis tasks require high throughput (Allcock, et al., 2002), running repeatedly from top to bottom through a very large unchanging or gradually growing data set and extracting meaningful information for interpretation.

The growing application of supercomputing for data analysis is not limited to fast detectors. It emerges in places where large volumes of data are collected, including the analysis of log files from web services (Mavridis & Karatza, 2017), large collections of high-resolution images or videos (El-Haija, et al., 2016), high-density sensor data (Lee, Gasster, Plaza, Chang, & Huang, 2011), and genomics (O'Driscoll, Daugelaite, & Sleator, 2013). Many of these tasks are data-parallel, meaning that subsets of the input data can be analyzed in parallel, reduced to a much smaller partial output, and the small partial outputs then combined to produce a complete result. This approach is known as the the "map-reduce" programming model (Dean & Ghemawat, 2008).  In general, meaningful reduction or filtering is always necessary to make large data volumes accessible for human beings.

Well-known frameworks that are based on the map-reduce design principle, such as Apache Hadoop and Apache Spark (Mavridis & Karatza, 2017), were originally established to analyze tabular data like log files and databases. The use of such approaches to handle n-dimensional binary data arrays such as high-resolution detector data is emerging for larger sizes of such data sets (Stegmaier, 2017).

In contrast to traditional high-performance computing or big data analytics, a data analysis system for interactive microscopy should handle live data while it is arriving from the detector and show a preprocessed monitoring result in real time, while at the same time saving raw data to storage. Furthermore, users often require interactive analysis and exploration with immediate *visual* feedback, ideally while a data set is still being acquired. Just as for previous smaller data sets, traditional file workflows such as saving, opening, browsing and copying should work in a familiar way, the only difference being the need to handle TBs on a remote distributed system instead of MBs on a local PC. Such systems must therefore be optimized more for low-latency high throughput than for full utilization of all computing resources as in traditional high-performance computing  (Goscinski, et al., 2014).

## Solution strategies

The computer scientist Donald E. Knuth is famously quoted for the phrase "premature optimization is the root of all evil" (Knuth, 1974). This phrase is often used to justify building a simple solution first and then optimizing it later. However, throughput in the TB/s region is out of the reach of an ordinary single PC. A system capable of such performance levels is difficult to develop incrementally from a simple PC-based solution and should therefore be designed from the start as a distributed, scalable high-performance solution. Such a system is inherently not the simplest one and it requires substantial optimization. This statement does not contradict Knuth, because his famous quote must be understood in its full context (Knuth, 1974):

> *Experience has shown … that most of the running time in non-IO-bound programs is concentrated in about 3 % of the source text. We often see a short inner loop whose speed governs the overall program speed to a remarkable degree; speeding up the inner loop by 10 % speeds up everything by almost 10 %. And if the inner loop has 10 instructions, a moment's thought will usually cut it to 9 or fewer.*

> *In established engineering disciplines a 12 % improvement, easily obtained, is never considered marginal; and I believe the same viewpoint should prevail in software engineering.*

> *We should forget about small efficiencies, say about 97% of the time: premature optimization is the root of all evil. Yet we should not pass up our opportunities in that critical 3 %.*

In order to achieve optimal performance, two factors can be maximized: A system should scale on many parallel worker units that perform the processing, and each of these workers should reach a high performance. The architecture should be validated and its critical code portions should already be identified and optimized in the system design phase, for example using prototypes. The following sections describe various aspects and design considerations for such a system.

## Optimization of processing performance

The throughput per worker can be optimized by identifying and resolving bottlenecks. Profiling yields detailed information on the amount of time the system spends in specific code portions and operating system functions. Based on this information, performance optimizations can be targeted at these critical parts of the system (Knuth, 1974). Array programming (Walt, Colbert, & Varoquaux, 2011), vectorizing operations (Larsen & Amarasinghe, 2000), eliminating unnecessary copies in the flow of data (Thadani & Khalidi, 1995) (Stancevic, 2003) and improving cache efficiency (Kelly, 2003) are common strategies that can be used to yield substantial improvements, especially when they are combined.

A scripting language such as Python is usually orders of magnitude slower than a compiled language, and using it for a high-performance application may seem counterintuitive. However, most of the code in a piece of software is not performance-critical (Knuth, 1974) and should be written in a language that has a low development and maintenance cost. Furthermore, most applications today use components of pre-existing third-party code and require a high-level glue language to connect them. Python is currently widely used for this purpose in scientific software because it combines easy development with excellent input-output (IO) and interfacing capabilities. Furthermore, it can achieve good numerical performance through suitable extensions. Numpy, for example, efficiently delegates standard numerical tasks to high-performance numerics libraries such as BLAS (BLAS (Basic Linear Algebra Subprograms), 2017) implementations, PyTorch (PyTorch, 2018) and Intel MKL (Intel, 2018) through array programming. Numba (Numba: A high-performance Python compiler, 2018) allows to compile selected code portions just-in-time to generate optimized machine code that can nearly reach the performance of equivalent code written in C or Fortran.

If no pre-existing optimized implementations of a processing task are available, then profilers and tools such as the Compiler Explorer (Godbolt) can be used to identify and take advantage of possibilities for speed-up.

## Optimization of IO performance and data layout

Fundamental design decisions hinge on whether a task is fundamentally CPU-bound or limited by IO. This point is not always easy to judge before optimization, because an optimized implementation can be more than 100 times faster than a naïve version (Data analysis unit, European Synchrotron Radiation Facility, Grenoble, 2018), which can make a CPU-bound implementation become IO-bound. The transfer of data between memory and CPU is often the ultimate bottleneck for numerical codes, not the computation itself (Alted, 2010). Investment in faster processing units or a higher number of cores per system does not necessarily increase performance. Distribution of the processing over many nodes, each of which has high IO and memory bandwidth, but with comparatively low processing speed and good energy-efficiency, can then be a more cost-efficient path.

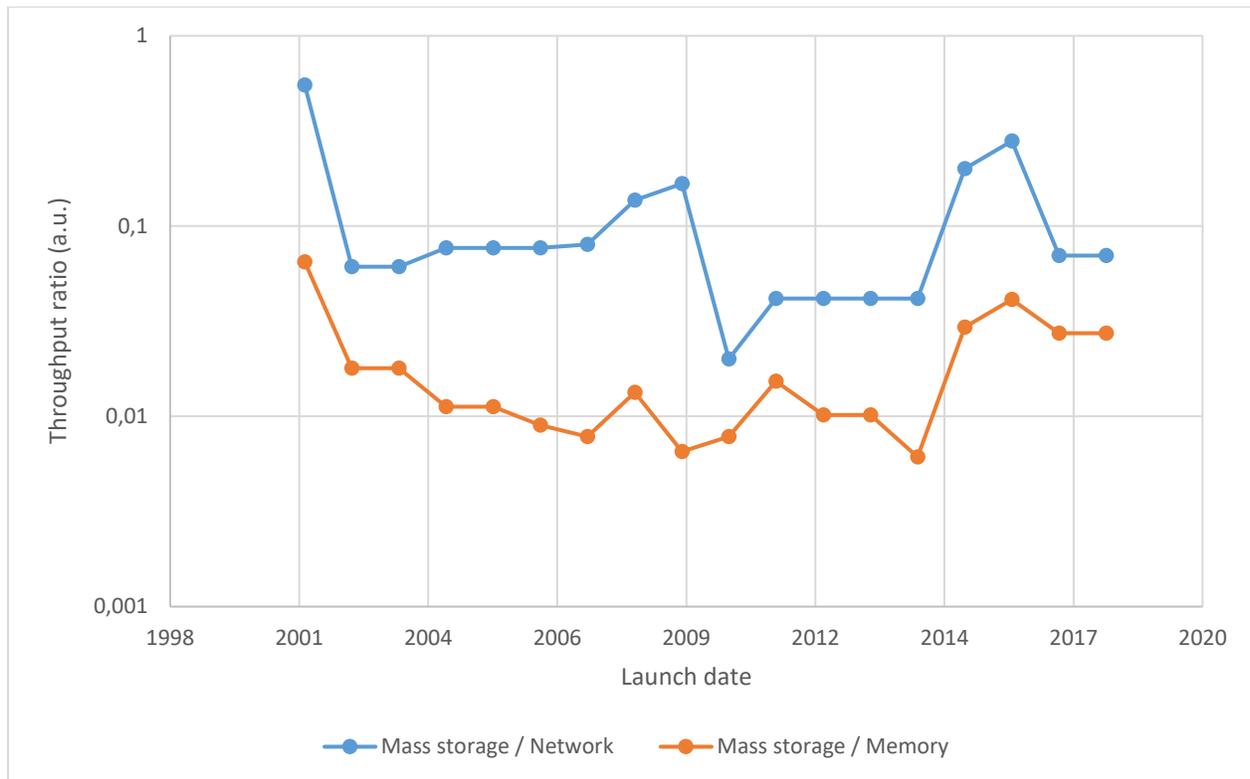

FIGURE 3: EVOLUTION OF MASS STORAGE (SSD AND HDD) THROUGHPUT RELATIVE TO NETWORK AND MEMORY BASED ON THE DATA IN FIGURE 1 (WEBER, 2018). THE MASS STORAGE THROUGHPUT IS DIVIDED BY THE NETWORK AND MEMORY THROUGHPUT TO HIGHLIGHT RELATIVE CHANGES. IT SHOWS A DYNAMIC DEVELOPMENT WITH FAST NETWORK CONNECTIONS EMERGING BETWEEN 2009 AND 2014, AND A DRAMATIC INCREASE OF MASS STORAGE PERFORMANCE AFTER 2014 DRIVEN BY INNOVATIONS IN THE SSD MARKET. RECENTLY INTRODUCED FAST NETWORKING STANDARDS LIKE 100 GBIT ETHERNET AND 400 GBIT ETHERNET MAY AGAIN CHANGE THE THROUGHPUT RATIO IF THEY FIND MORE WIDESPREAD APPLICATION.

High IO rates to make data available for processing at worker nodes can be achieved using high-performance network connections to dedicated storage nodes, or by using fast local storage on the worker nodes themselves. Each method can have advantages and disadvantages, depending on the application and available components. As a result of a dynamic development in IT between the years 2000 and 2018, the boundary conditions for designing such systems have changed considerably over time, resulting in a profound impact on the design of high-performance computing infrastructure.

The read rate of traditional hard disk drives (HDDs) fell behind the increase of network performance between 2010 and 2014 (Figure 3). As a result of relatively slow mass storage before 2014 compared to the available fast network interfaces (Figure 3), it then made sense to build systems that were based on dedicated storage nodes with large disk arrays (RAIDs) of relatively slow magnetic disks with fast network connectivity to worker nodes. This used to be the most attractive solution for high IO performance at computation nodes (Intel High Performance Data Division, 2014).

Between 2014 and 2018, flash-memory-based solid-state disks (SSDs) that are directly connected to the ultrafast PCIe buses of modern computers (NVMe SSDs) have provided tremendous progress in performance, while their cost has dropped over time. NVMe SSDs with a capacity of 2 TB that allow reading at around 3.5 GB/s are affordable stock components in 2019. These ultrafast SSDs have closed the gap between high-speed network connections and mass storage, making local storage competitive when compared to network-based solutions.

For all tasks that are data-parallel, i.e. for which the input data can be split into chunks that are stored on separate nodes and processed repeatedly with the map-reduce pattern as described in Section "Challenges and requirements", such a setup can be significantly more cost-effective per aggregate throughput than a network-based solution. It eliminates the need for high-speed storage nodes and only requires commodity network connections such as 10 Gbit Ethernet to initially populate the SSDs, send queries and transfer the results, so long as the results are smaller than the input data (Patel, Birla, & Nair, 2012). The drawback is that only data-parallel tasks, where each worker reads or writes a small pre-defined part of a data set, benefit from such an architecture, while a network-based setup with a suitable cluster file system allows fast random reads and writes over all regions of a data set.

A cluster that is optimized with fast local storage, but has slow connections between nodes, will only perform well for workloads that involve fast data-parallel reading, processing and writing, and not for workloads that require frequent high-volume low-latency data exchange between nodes, such as running large-scale iterative simulations or frequently re-shuffling a data set.

The Hadoop File System (Hadoop FS) (Shvachko, Kuang, Radia, & Chansler, 2010) was developed specifically to manage distributed local storage on processing nodes. It provides an interface of one coherent file system over all storage nodes, while managing splitting into blocks, transfer and replication in a way that is transparent to the user. In contrast to traditional distributed file systems such as Lustre (Lustre Software Release 2.x Operations Manual, 2018), Hadoop FS is not POSIX-compatible, that means it does not provide all the functionality of a typical file system that conforms to the POSIX standard. The primary difference is that it is append-only, i.e. it does not allow random writes to an existing file. Since it is optimized for workloads that stream a large data set to storage from top to bottom and then read many times from this unchanging source file, this is not an obstacle and removes the synchronization effort that would be required to manage changes in a distributed system. A naïve application can read from the Hadoop FS just as from a cluster file system, accessing all parts of a file. This access method typically requires data transfer between the nodes which provides acceptable performance if the network is fast enough – as for a traditional cluster file system.

An application for data-parallel tasks that is aware of the Hadoop file system, can query it about the partitioning and distribution of blocks, partition the processing in a compatible schema, and match tasks to the nodes that hold the required data on their local storage. Under these circumstances, the worker can perform a so-called "short-circuit read" (Apache Software Foundation, 2018) whereby the application obtains an operating system file handle directly to a data block on the local storage medium.

The application can access this block directly through local operating system routines with zero overhead, including memory-mapping, through that handle. In such a setup, the full data is sent through the network only once for writing to local storage on the nodes. Reading is limited to local transfers from the SSDs and does not strain the network.

Depending on the nature of the processing routines, it can be advantageous to re-order the data in order to bring different parts that have to be processed in combination close together on the storage device. An example comes from the use of tiled detectors. One would most naturally write separate files for each tile to achieve high acquisition throughput. The data for a complete frame is then always split up over separate files that could be stored on separate nodes to achieve the highest possible acquisition throughput. However, if a Fourier transform of an entire detector frames is required, then it can make sense to re-shuffle the data so that complete frames are stored together on the same node, as combining the Fourier transforms of partial frames requires substantial data transfer (Cooley & Tukey, 1965). If all of the data from one frame are kept on the same node, then the data does not have to be sent across the network.

## Distribution of computation

Systems such Apache Spark (Zaharia, et al., 2016), which is written in Scala, as well as the Python-based Dask and Dask.distributed (Rocklin, 2015), were developed to make computation on distributed systems easier to manage than using the message passing interface (MPI). Just as the Hadoop file system manages the distribution of storage blocks, Spark and Dask manage the execution flow of computation by processing and combining smaller portions as needed. Spark is optimized to work with distributed storage on the Hadoop file system which is developed specifically for such tasks. It is applied mostly to tabular data, with few provisions for large-scale numerical calculations at the time of writing. Dask handles both tabular and numerical data and has access to the PyData ecosystem, a rich interoperable collection of open source Python modules for a wide range of numerical processing and visualization tasks. Unfortunately, it has no native support to reap the benefits of distributed local storage following the principle of the Hadoop file system at the time of writing.

## Graphical user interface

In contrast to typical graphical user interface (GUI) applications, if data processing does not take place on the same PC but on a distributed system, then the front-end and back-end can be far apart, with potentially large round-trip times, over unreliable connections, and with limited throughput between the front-end and back-end. Operations that would finish instantaneously and reliably within a traditional GUI application that runs on a single PC may then take much longer or fail when using such a distributed setup, for example by losing network connectivity between the front-end and back-end. This situation makes asynchronous operation critical and introduces many more transition and error states that need to be handled.

The development of such a hybrid solution for an interactive GUI with a distributed high-performance data handling and processing back-end requires a suitable system architecture and design approach (Foust, Järvi, & Parent, 2015) (Abramov, Redux: Core Concepts, 2018) (Abramov, The Case for Flux, 2015) (Taylor, et al., 1996) (Rogic), which is different from previously known batch processing systems that are used in large-scale scientific computing. Large social networks such as Facebook arguable come closest to fulfilling similar requirements (Shneiderman, Plaisant, Cohen, & Jacobs, 2009), as they have a remote distributed back-end, deal with large-scale data and have a responsive local GUI. On the scientific side, IT systems of the XFEL (Fangohr, et al., 2018) or systems such as the cesium platform for time-series inference (Naul, Walt, Crellin-Quick, Bloom, & Pérez, 2016) fit into this category.

The React framework for web applications (Facebook Inc., 2018) provides a rich ecosystem for user interface components, in combination with state management through libraries such as Redux (Abramov, Redux: A predictable state container for JavaScript apps., 2018) (Abramov, Redux: Core Concepts, 2018) (Abramov, Redux: Usage with React, 2018) and Flux (Abramov, The Case for Flux, 2015). These frameworks were developed specifically to manage the complex state and asynchronous operation of GUIs for distributed systems. A web-based GUI allows a system to be accessed using standard web technology through any device that can run a modern web browser, making it a simple and universal choice to develop a responsive GUI for distributed data analysis tasks. Using traditional GUI frameworks is also possible, but requires more work to enable asynchronous operation and to manage the state.

### IT infrastructure management

The development and deployment of a distributed system requires coordination of the operation of many separate computers, including installation, deployment and network configuration. Furthermore, failures of parts of the system occur more frequently when there are more components and higher complexity.

Container management solutions such as Docker (Merkel, 2014) and Kubernetes (Bernstein, 2014) can help to orchestrate the many components of a distributed system, including deployment on commercial or scientific cloud infrastructure.

### People

Programming languages and software libraries have become so easy to use that basic programming skills are often sufficient to implement effective data analysis workflows and test novel processing schemes with acceptable performance (Schindelin, et al., 2012) for data volumes and computations that are compatible with PCs.

In contrast, the assembly of existing infrastructure into an effective and user-friendly distributed high-performance processing system for large volumes of data requires a deep understanding of software design and IT systems, from CPU and GPU architecture and instruction sets, through memory and IO handling on the hardware and operating system level, to network architecture, network protocols and web development for remote GUIs.

For this reason, more people who have a primary background in computation and system design are required in this emerging field (Shaw, 1990). Furthermore, data analysis is becoming an interdisciplinary topic between computer science, image processing and application science. In this environment, the development of innovative high-performance IT systems is becoming a key enabler to support progress in the scientific analysis and exploitation of such large-scale data.

### File format requirements

Advanced acquisition and processing tools allow users to generate more data and more complex data sets, and to perform more complex analysis. Furthermore, the movement towards Open Data and Open Science means that users not only exchange analysis results but also publish original data, together with corresponding analysis tools.

For this reason, a standardized open format with rich and reliable metadata that give meaning and context to a data set and makes it interoperable, discoverable and searchable, is required (Zuiderwijk, Jeffery, & Janssen, 2012). Raw data, i.e. a huge collection of data values, is only interpretable in combination with additional information from metadata. Convenient analysis tools should be able to display and meaningfully interpret metadata, ranging from basic information such as the number format and array layout to the instrument settings and sample condition during acquisition.

The file format for detector data should be carefully designed. For raw data, it should allow very efficient, distributed, parallel reading and writing from and to distributed storage, such as the Hadoop file system. For correct and easy interpretation, it should include metadata that follows a standardized schema.

At the time of writing, there is no single catch-all solution. The HDF5 format (The HDF Group, 2016) provides a container that fulfills the requirements with regards to metadata, but it is unfortunately, in its basic Open Source version, incompatible with the way in which systems like the Hadoop FS reach high aggregate IO performance on distributed storage. Furthermore, HDF5 is currently in the process of moving to an "open core" business model, where advanced features, such as support for Hadoop FS, are Closed Source (The HDF Group, 2018), making it unattractive as a dependency for Open Source high-performance solutions. The use of raw binary data blocks that can be distributed across a distributed storage solution and accessed with the full range of optimized file reading methods that the operating system provides on processing nodes, such as memory-mapping or direct IO, in combination with a sidecar file that contains metadata in a standardized format, currently provides a workable solution that achieves very good IO performance on distributed storage. Such a data layout can be created from a source file through an ingest step or potentially using a transparent caching solution.

## Open Source requirement

The publication of such a system under an Open Source license is the only reasonable licensing model (Willinsky, 2005).

First, it is scientific software, which means that data processing should be transparent for review. The source code is a direct human-readable description of what a given software does. Scientists as users should be able to modify the software in order to perform new types of analysis. Since a future analysis method can be fundamentally new and different from anything anticipated before, every aspect of the software should be modifiable so that it can be adapted and repurposed for new applications, and to take advantage of new developments in high-performance IT infrastructure.

Second, development is often directly or indirectly publicly funded with tax revenues. Consequently, the results should be available to all stakeholders under fair, reasonable, and non-discriminatory (FRAND) terms, just as the publication of scientific results. Open Source licensing is a straightforward and widespread licensing model to satisfy this requirement.

Third, high-performance software is recompiled often in order to reach optimal performance on a given setup, requiring access to the source code.

Fourth, installation and cluster deployment are easiest in the framework of an Open Source licensing model. License and compliance management are then straightforward both for developers and for users of such distributed software. There are no technical and very few legal prerequisites to deploy and use the software in any desired form. This situation is very much the opposite when using Closed Source software.

Fifth, Open Source code is advantageous as a foundation for other people's work, for example for integration into third-party software. The use of an Open Source licensing model means that users can rely on the solution remaining available and serviceable under predictable terms and conditions. Even if the original developers abandon the software, the license allows third parties to step in and continue maintenance and development.

Sustaining such a project requires other contributions than income from license sales (Okoli & Nguyen, 2015). Most notably, an Open Source project benefits from in-kind contributions of code, suggestions for improvement, documentation, marketing and support, thereby blurring the lines between developers, managers and users. Other sources of support can be public funding if the software or specific developments are in the general interest of the funding body or enable a project. Support through foundations such as NumFOCUS (NumFOCUS, 2018), the Apache foundation (The APACHE Software Foundation, 2018) or the Linux foundation (The Linux Foundation, 2018) plays a big role for key infrastructure projects. Companies or institutes that are heavy users of a software package can employ key personnel that dedicate most of their time to its development and benefit directly from improvements, maintenance and support. Technology companies like Microsoft (Microsoft, 2018), Google (Google, 2018) and Facebook (Facebook, 2018) even develop many in-house projects that primarily serve their own business interests openly as Open Source in order to benefit from additional contributions from third parties and to improve the general user and developer experience for their customers at no cost. Further sources of income can be donations, and service contracts for support and maintenance.

## LiberTEM as an example

LiberTEM is an Open Source platform for high-throughput distributed processing of pixelated STEM (Williams & Carter, 1996) data, which is being developed in response to challenges and opportunities in handling big data sets from fast detectors for electron microscopy. This section describes the design considerations that guided its initial development in 2018, against a backdrop of existing technology and design principles.

The primary focus of the LiberTEM platform is currently the handling and analysis of pixelated STEM data. The most basic operation for this application is to emulate traditional STEM detectors, which integrate part of the scattered intensity. The pixels of each detector frame are multiplied element-wise with a set of masks and the products for each mask are added up for this type of analysis. A simple binary mask allows the result from any standard STEM detector to be replicated and virtual darkfield imaging to be performed. More complex combinations of masks that have gradients and negative values allow, for example, center of mass calculations and background subtraction from diffraction peaks.

Advanced processing beyond the use of masks can involve any form of image analysis and mathematical processing. Pixelated STEM allows the re-analysis of data using different processing methods on a computer, instead of re-acquiring data at the instrument using different settings, and it allows analysis schemes to be implemented that would be difficult or impossible to achieve with physical systems.

The basic mathematical operation of applying a stack of masks can be simplified by flattening (vectorizing) the scan dimension of a data set into a linear list. The usually rectangular detector and mask data can be flattened to vectors. The operation becomes a matrix-matrix multiplication of a frame stack with a mask stack, which is one of the most optimized and well-known data processing tasks (Goto & Geijn, 2008). At the same time, it is "embarrassingly parallel". All frames and masks are independent, all element-wise multiplications of a detector pixel with a given mask element are independent, and the summation of these products is associative. The operations can therefore be chunked in nearly any way, the result can be easily assembled from partial results in different ways, and the result is very compact when compared to the input data, since the number of masks is usually low in comparison to the number of detector pixels and frames. It is the ideal situation for processing the data using the map-reduce scheme.

Since modern CPUs and GPUs can perform matrix-matrix multiplications very quickly, the primary task of the LiberTEM platform is to distribute the IO and to keep many cores fed with data, while providing an easy-to-use interface, both in terms of GUI and scripting. No single distributed processing framework currently fulfills all these requirements for interactive pixelated STEM data processing (Mehta, et al., 2016). For this reason, LiberTEM is assembled from several different components.

In order to make the transition from processing on PCs to distributed systems easier, LiberTEM is designed to be easy to install and run on any PC. Unless it is configured otherwise, running it using default parameters starts a local pool of workers, with all components on the same computer. The ambition and value proposition of LiberTEM is to match or exceed the single-node performance and usability of any other processing solution for the supported applications, making it valuable not only as a cluster solution but also as a single node system.

## Spark compared to Dask.distributed

Both Apache Spark and Dask.distributed are designed to make distributed computation easier. In principle they perform very similar tasks (Anaconda Inc., 2018), as described below.

On a fundamental level, Spark comprises a well-designed combination of distributed data type, language, execution engine and compiler, with excellent support for data locality on the Hadoop FS. It is written in Scala, which is a functional language that is compiled to run on the Java Virtual Machine (JVM). The primary bottleneck of Spark, when the fundamental design decisions for LiberTEM were made at the beginning of 2018, was the IO performance of the JVM for numerical. When Spark reads binary numerical data from disk and passes it on *via* the Java Native Interface (JNI) to external libraries like a BLAS implementation, the JVM first reads into an input buffer, copies the data from the input buffer into an internal buffer, and then copies it again into another buffer, which it then passes *via* pointer to the external library. This process limited throughput on a test system to 2.7 GB/s for cached file system reads and typical LiberTEM processing routines. Furthermore, the ecosystem for scientific numerical processing in Spark is much less developed than the Python world at the time of writing.

In contrast, memory mapping a file or letting the operating system copy data into a buffer to pass it on to an external library works with very little overhead in Python. The same task on the same system reaches 14 GB/s, which is more than five times the performance of Spark, with a system implemented in Python that uses Dask.distributed. For this reason, this framework was chosen instead of Spark to distribute tasks on workers.

Unfortunately, Dask.distributed had no native support to exploit data locality when reading from the Hadoop FS at the time of writing. We implemented chunked processing with so-called Dask futures. Such a future defines a processing task, in the context of LiberTEM typically processing a specific part of a data set, in the form of a function call that the Dask scheduler dispatches for asynchronous execution on one of the workers. The Dask scheduler later makes the result of this function call available to the process that issued the future. This allows us to implement support for data locality ourselves by issuing futures that are assigned to specific workers based on the data layout, and gives us more direct control about the execution process for the live updating of results and for the general subdivision of processing jobs. Furthermore, we can handle live data streams by adding processing jobs for chunks of data as they are coming in.

Dask.array, an emulation of a global data array that is distributed over the processing nodes, is a popular interface to define distributed numerical processing workflows with Dask. Processing tasks are defined in terms of operations on this global array, and Dask computes a task graph to perform the processing task with smaller step-wise calculations on partial data in a distributed and parallel fashion. The task graph is pre-calculated before computation starts and can reach a significant size if a large data set is processed in small chunks. Furthermore, the task graph calculated by Dask is not always the optimal solution for a given processing task, and it can be difficult to understand and influence how a specific processing task is translated into a task graph. Calculating such task graphs in an optimal fashion for distributed systems that perform non-trivial operations, taking into account data locality and efficiency on all levels, is a hard problem and currently the subject of advanced compiler development (Grosser, et al.).

The flexibility and simplification of calculating task graphs for arbitrary operations is not required in our case since we perform highly optimized recurrent tasks, for which the sequence of processing steps can be hard-coded as the execution flow of our software. With Dask futures, we work on small chunks with an explicitly defined processing sequence that optimizes efficiency on all levels, without taking up memory and computation time for a fine-grained complete task graph.

Benchmarking results for LiberTEM are presented below in the section "Benchmarking results".

## Architecture

The LiberTEM architecture, which is shown in Figure 4, is built to support high-throughput map-reduce-style computations on big data in a streaming fashion.

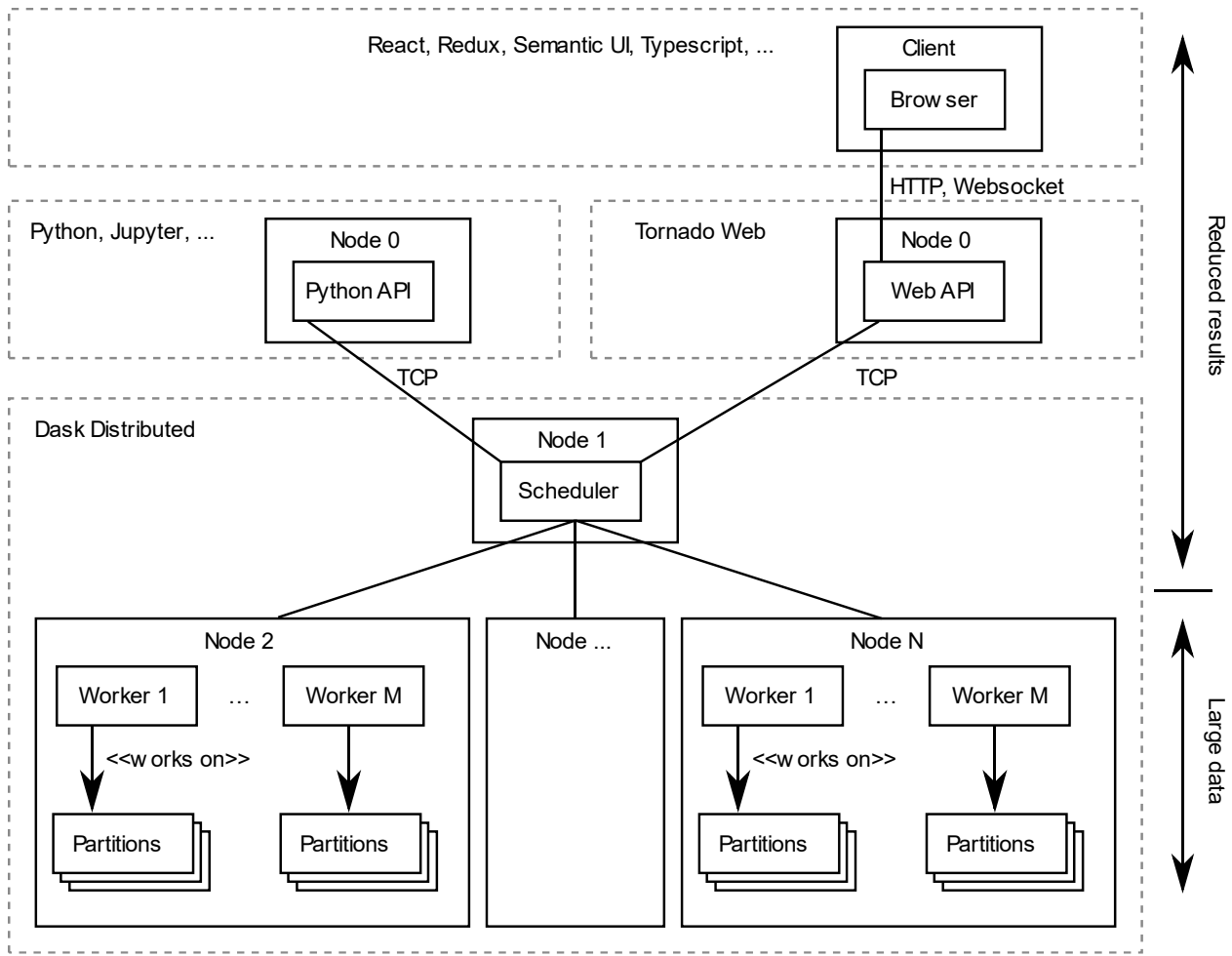

**FIGURE 4: ARCHITECTURE OF LIBERTEM TO PERFORM MAP-REDUCE OPERATIONS ON LARGE SCALE BINARY DATA SETS WITH HIGH AGGREGATE THROUGHPUT. A DATA SET SPLIT INTO PARTITIONS THAT ARE ASSIGNED TO THE WORKERS ON THE PROCESSING NODES. THAT ALLOWS TO STORE THE PARTITIONS ON LOCAL STORAGE FOR FAST AND SCALABLE READ ACCESS. THE WEB API OR PYTHON API SENDS PROCESSING TASKS THAT OPERATE ON THESE PARTITIONS TO THE SCHEDULER, WHICH THEN DISPATCHES THEM TO THE WORKERS AND FEEDS THE RESULTS BACK TO THE API. THERE, THE PARTIAL RESULTS ARE ASSEMBLED AND RETURNED TO THE USER. THIS PROCESS CAN BE PERFORMED ASYNCHRONOUSLY AND PROGRESSIVELY TO ACHIEVE LIVE UPDATING RESULTS ON THE USER SIDE.**

The high-level part of LiberTEM that runs on the control node deals with job handling, tasks, division of data and work, and scheduling. This part does not touch the source data, but only handles abstract descriptions of the data, as well as the processing results that are reduced in size. It consists of a client GUI running in the user's browser, which communicates with a web application programming interface (API) *via* HTTP and WebSockets. The web API connects to a Dask.distributed scheduler, which is then responsible for distributing the work onto worker processes and across machines.

As an alternative to the GUI, the user can connect directly to the Dask.distributed cluster *via* the Python API, for example from a Jupyter notebook, from software written in Python, and from software that includes an embedded Python interpreter. An API for other programming languages is currently in planning.

The low-level part of the architecture handles the source data. It consists of worker processes that are distributed on all machines. The worker processes execute tasks, which operate on partitions of the source data. The partitions are statically assigned to nodes, for example using a Hadoop FS namenode. The partitions can be logical blocks of one or several larger files, or they can correspond to individual files of a multi-file dataset. Partitions that are composed of several smaller files are also possible.

The system operates in an asynchronous fashion for progressively updating processing results on the user side. The Python API can wrap this into a batch-oriented synchronous workflow, for example for scripting.

## Execution and data flow

In order to start processing a data set, the user creates a job, which is an abstract description of the operation to perform, including a reference to a dataset with the parameters for the operation, for examples the masks to apply.

The user can then instruct LiberTEM to run the job. The dataset is subdivided into partitions, whose size is chosen in such a way that their processing finishes in about 100 ms to achieve a real-time feeling for the user (Nielsen, 1994). A size of 256 MB to 512 MB achieves this response time for typical mask application tasks when reading from the file system cache. Smaller partitions create more overhead for scheduling, while larger partitions result in less responsive feedback.

For each partition, a task is created, which is scheduled by Dask.distributed. It is executed on a worker that has fast access to the partition, i.e., one running on a node that keeps the partition on its local storage. The high-level task handling part of LiberTEM can query a Hadoop FS name node to list the nodes that hold a given partition. Details about task execution within Dask.distributed are given elsewhere (Anaconda, Inc, 2018).

When a task is executed on a worker process, it reads source data from its associated partition. This is done preferentially by memory mapping (mmap) the entire dataset or partition and delegating cache-efficient access to the underlying computation if the node has enough memory to keep all source data in its file system cache. For formats that require more pre-processing or decoding before computation, or under memory pressure, the partition is read in small tiles that fit comfortably into the L3 cache for each worker process. We currently run one worker process per CPU core that processes a single tile at a time. Tile sizes of approximately 1 MB give good results for the typical L3 cache sizes of current CPUs.

After running the computation, the reduced result is sent back to the user's machine and combined with the results of the other tasks to build up the complete result.

When using the web API and web-based GUI, the job is created when the user clicks the 'Apply' button. As a job can take some time to complete and the web API may need to respond to other requests in the meantime, this part of the code is written using the Asyncio package and async/await keywords that were introduced in Python 3.5. In this way, the event loop of the web API is never blocked during long-running operations and remains responsive.

## GUI and API

Figure 5 shows a screenshot of the web application GUI of LiberTEM. Most user interaction, such as changing the size and position of the mask or adding a new analysis, runs locally in the browser. Clicking "Apply" starts a calculation on the back-end, i.e. creates a job and sends processing tasks to the workers.

The GUI remains responsive while an analysis is running and shows partial results as they arrive from the processing back-end. The GUI allows several windows, files and analyses, which can operate in parallel, to be opened.

Figure 6 shows an example of using LiberTEM through its scripting API in a Jupyter notebook. The asynchronous back-end is encapsulated with a synchronous wrapper to make it easy to use in such a scripting environment. The asynchronous API can be used for integration in a third-party application with the same responsive and interactive behavior as the web GUI.

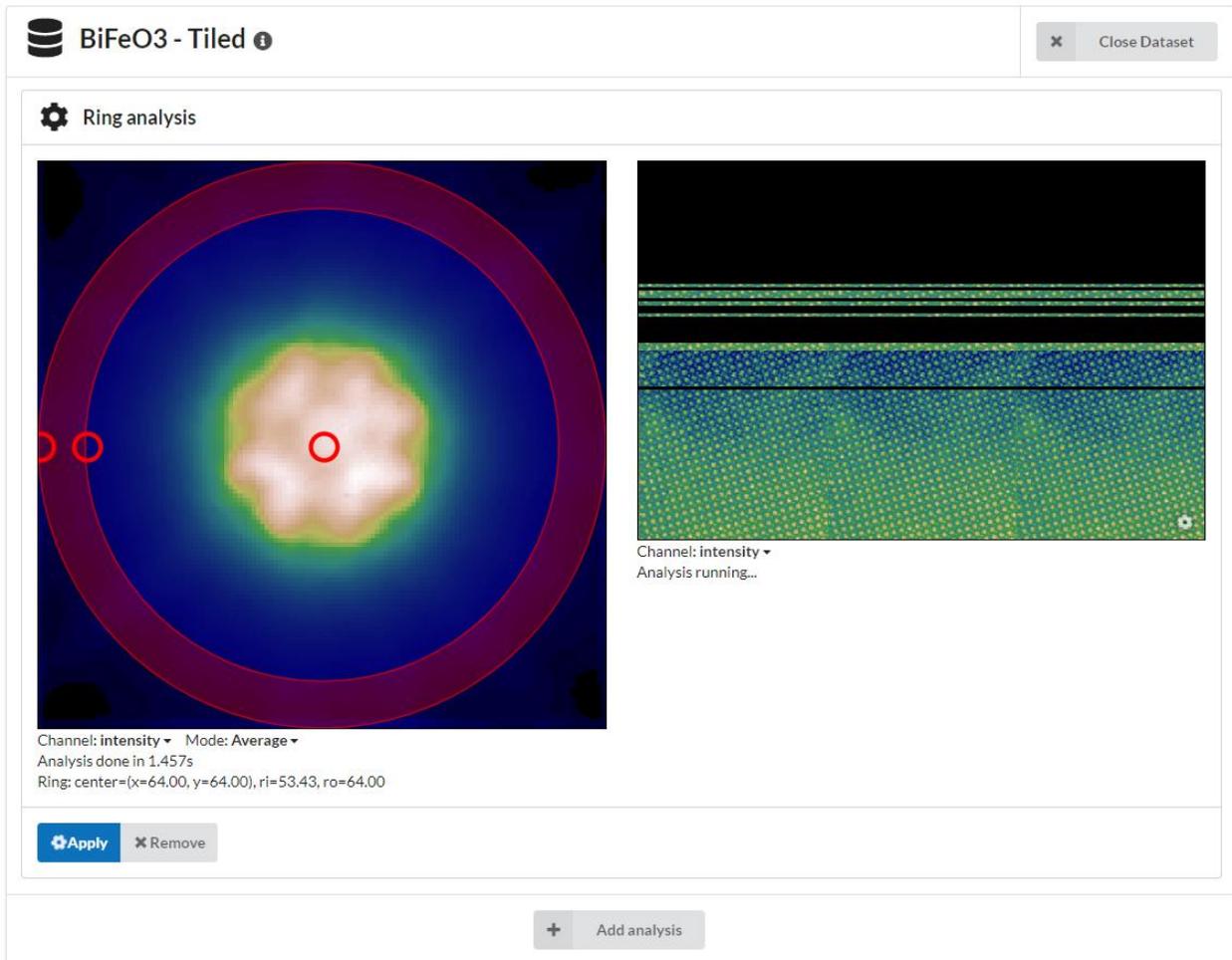

FIGURE 5: GUI OF LIBERTEM WHILE PROCESSING A VIRTUAL HIGH-ANGLE ANNULAR DARKFIELD IMAGE ON A 24 GIB SEMI-SYNTHETIC 4D STEM DATASET. ON THE LEFT, THE GUI CAN SHOW AN AVERAGE OF ALL DETECTOR FRAMES OR INDIVIDUAL FRAMES. THE USER CAN DEFINE THE SIZE AND POSITION OF THE RING THAT IS USED AS A MASK FOR INTEGRATING THE INTENSITY ON ALL DETECTOR FRAMES. ON THE RIGHT, THE PROCESSING RESULT IS DISPLAYED IN THE FORM OF A VIRTUAL STEM IMAGE. THE GUI SHOWS PARTIAL RESULTS WHILE THE CALCULATION IS PROGRESSING. EACH STRIPE CORRESPONDS TO AN INDIVIDUAL PARTITION. BLACK STRIPES INDICATE RESULTS THAT ARE STILL PENDING.

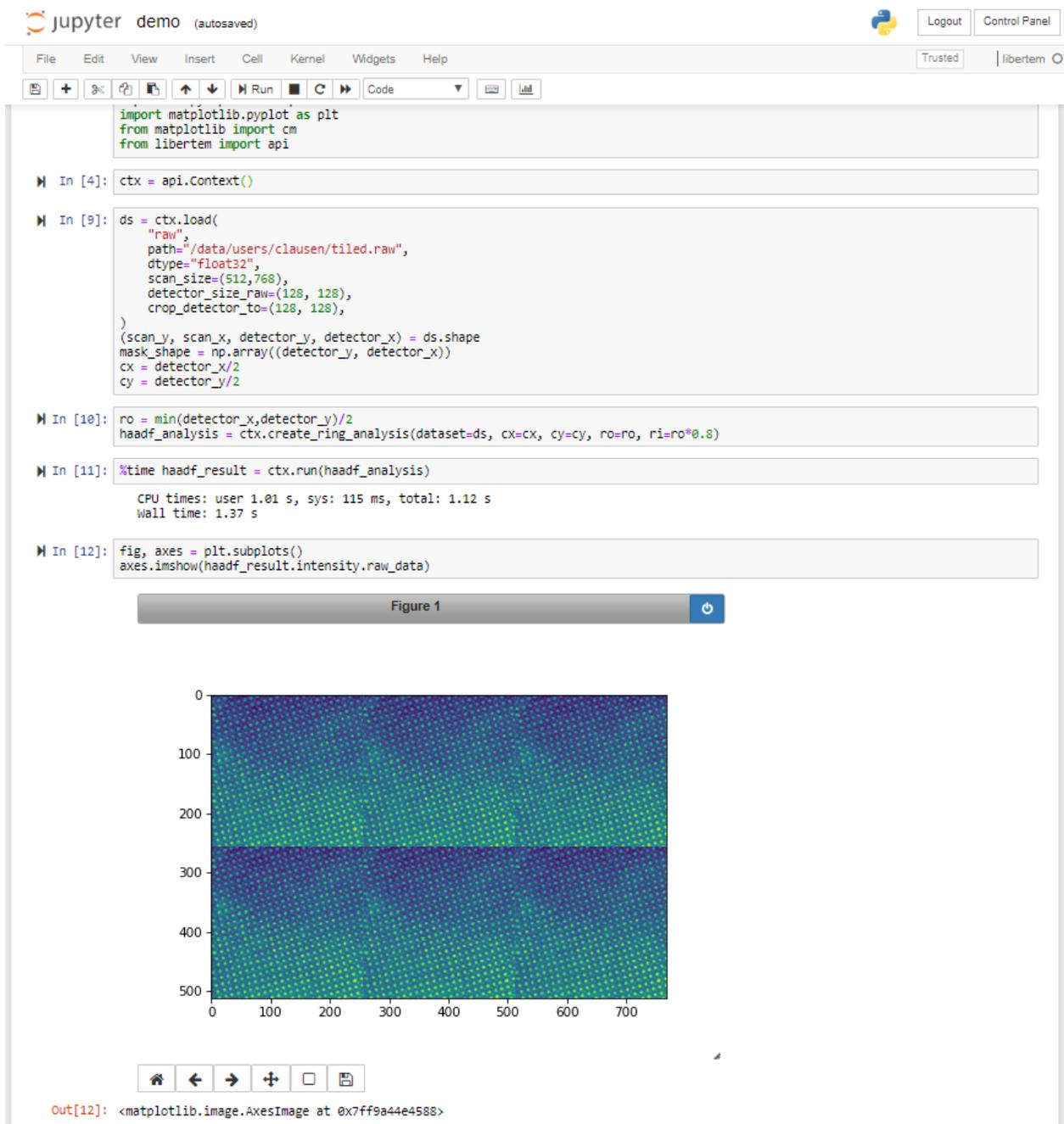

FIGURE 6: EXAMPLE OF USING LIBERTEM WITHIN A JUPYTER NOTEBOOK. THIS SCRIPT CALCULATES A SIMILAR RESULT AS IN FIGURE 5 USING THE PYTHON API OF LIBERTEM. DETAILS REGARDING THE API AND FURTHER EXAMPLES CAN BE FOUND ONLINE AS PART OF THE LIBERTEM DOCUMENTATION.

## Numerical processing

Since LiberTEM implements basic pixelated STEM processing using matrix multiplication, it takes advantage of existing high-performance software libraries for this purpose. Usually, only a few masks are applied per processing job, meaning that the mask stack matrix is tall and skinny. The data set can be non-contiguous when parts of it, such as frame headers, are masked out. Under these circumstances PyTorch currently achieves the best performance, up to two times faster than Numpy using Intel MKL

BLAS or OpenBLAS. As a fallback, LiberTEM can use the BLAS implementation that is accessible through the Numpy Python package on the system.

More complicated processing schemes may require the development of efficient kernels, as a Python and Numpy implementation typically cannot use all possibilities for optimization, like variable elimination and optimal use of CPU caches and registers. Bit packing and unpacking is an example for a processing task that is difficult to implement efficiently using Numpy. Such kernels are often written in C, C++ or Fortran, because these languages traditionally have the best compilers and tools to transform the source code into highly efficient machine instructions.

Since only a very small percentage of the source code, perhaps even only a few lines, is actually performance-critical (Knuth, 1974), the use of a different programming language with its own full-blown build chain just for these lines of code is a disproportionate effort. Furthermore, compiling the code to use optimal machine instructions for any given machine can create significant efforts and introduce error sources both for development and deployment of a software project.

Numba, a package that compiles Python code of selected functions just-in-time (JIT) to highly optimized machine instructions for the native CPU or GPU architecture through the LLVM compiler infrastructure (The LLVM Foundation, 2019), has turned out to be very promising. When compared to an optimized C, C++ or Fortran implementation, Numba source code is much easier to manage and deploy on a given machine. Numba manages a self-contained small build chain that is very easy to deploy on a target system simply by installing the Numba Python package. When combined with the flexibility of Python, it even allows new versions of a function to be compiled on-demand, which can be advantageous under some circumstances.

For numerical processing, it reaches nearly the same performance as highly optimized C or C++ code, because it can use the capabilities of the LLVM stack, most notably variable elimination and auto-vectorization to use "single instruction, multiple data" (SIMD) instruction sets. Furthermore, it allows routines that process the data in fine-grained blocks to be written, which can benefit from the L1 and L2 CPU caches and would incur significant overheads for Python with Numpy.

As an example, LiberTEM uses a Numba implementation of a fast vectorized decoder for packed uint12 data (Figure 7) as part of a high-throughput file reader that reaches 85 % of the performance of an equivalent C version. Although this may sound trivial, bit packing and unpacking with optimal throughput using SIMD instructions can be a challenging task because of data alignment issues. Only modern compilers can generate efficient autovectorized machine instructions for this task from simple source code. Hand-optimized intrinsics with very significant development effort had to be used previously (Lemire & Boytsov, 2012). In contrast, the implementation in LiberTEM is *both* simple *and* fast.

```python
@numba.njit
def decode_uint12_le(inp, out):
    """
    decode bytes from bytestring ``inp`` as 12 bit into ``out``

    based partially on https://stackoverflow.com/a/45070947/540644
    """
    assert np.mod(len(inp), 3) == 0

    for i in range(len(inp) // 3):
        fst_uint8 = np.uint16(inp[i * 3])
        mid_uint8 = np.uint16(inp[i * 3 + 1])
        lst_uint8 = np.uint16(inp[i * 3 + 2])

        a = fst_uint8 | (mid_uint8 & 0x0F) << 8
        b = (mid_uint8 & 0xF0) >> 4 | lst_uint8 << 4
        out[i * 2] = a
        out[i * 2 + 1] = b
```

FIGURE 7: FAST 12 BIT DECODER WRITTEN IN PYTHON AND COMPILED WITH NUMBA. BY USING THIS DECODER, LIBERTEM REACHES AN AGGREGATE PROCESSING THROUGHPUT OF 8 GIB/S ON A XEON W-2195 WITH 18 CORES WHEN APPLYING A SINGLE MASK OPERATION TO K2 IS RAW FILES THAT CONSIST OF PACKED 12 BIT INTEGERS, CORRESPONDING TO 21 GIB/S OF FLOAT32 NUMBERS PIXEL-FOR-PIXEL, WHICH IS AS FAST AS PROCESSING FLOAT32 DIRECTLY USING LIBERTEM.

## Processing unit

In principle, graphics processing units (GPUs) are excellent at performing matrix multiplications. However, GPUs are diverse, not all PCs have one that can outperform its central processing unit (CPU) for mathematical tasks, and not all GPUs have fast enough memory transfer to match the speed of the CPU in throughput-limited tasks. For these reasons, LiberTEM currently implements a fast CPU version of all of its capabilities. The use of available GPUs as "boosters", with both CPU and GPU workers in parallel, is in planning at the time of writing.

## Support for the Hadoop FS

Since LiberTEM should run on any PC as a local system, it can read from any file system. For cluster deployment, a system like the Hadoop file system is ideally suited for a massively data-parallel task like pixelated STEM data processing. For this reason, LiberTEM is designed to take maximum advantage of such a setup if it is available. Not all file formats, most notably HDF5, are well suited for distributed storage. LiberTEM therefore implements an optimized custom data layout for distributed storage that can be created from any suitable input file using an ingest step. The layout consists of a sidecar file with metadata and several raw binary files that constitute the input data partitions used in processing. A distributed file system can then distribute and replicate these separate blocks through the cluster.

The processing follows this partitioning scheme and distributes tasks to the nodes that hold such a partition in their storage. The workers can use the short-circuit read process of the Hadoop FS for blocks that are on the local storage. In this way, LiberTEM achieves high aggregate throughput and excellent scalability.

The "Benchmarking results" that are given below were not generated on a Hadoop FS, but simulate such a setup by copying the entire data set to each node to simplify the tests.

## File format support

LiberTEM provides users with high throughput, responsiveness and ease of use, in particular for initially exploring a new data set after acquisition. It allows many file types to be read in order to avoid file conversion where possible. Current support includes raw data such as data from the EMPAD detector (Tate, et al., 2016), Quantum Detectors MIB, Gatan K2 IS raw data, Nanomegas Block Files, SER files, FRMS6 files for PNDetector, and all usual HDF5-based formats, such as NeXus, Hyperspy and EMD files.

For read performance, the ideal format is raw data in one of the standard number formats supported on the CPU architecture and the underlying BLAS library, most notably float32 in row order (C order), i.e., detector frame after detector frame without any frame headers or other gaps. Such a data set can be memory-mapped and passed directly to numerics libraries such as PyTorch or BLAS. Interaction between the numerics library's memory access pattern and the operating system's paging mechanisms for memory mapping provides optimal performance if enough free memory is available. If the file is already in the file system cache, then accessing a memory-mapped file on Linux is as fast as accessing a buffer in memory.

File formats that store packed integers, like uint12, are currently able to reach approximately the same performance level number-for-number. The processing time for the additional decoding step and transfer into a small extra buffer is compensated by a much smaller size on the storage medium, which improves the read rate as well as the memory use of the file system cache. Working on buffers that fit comfortably into the L3 CPU cache means that the subsequent processing step can access the buffer from the L3 cache instead of the main memory.

HDF5 achieves good performance on single-node setups. It can only be memory-mapped for the special case of contiguous data without chunking, which means that it nearly always incurs overheads from copying and HDF5-internal processing. On a Hadoop file system, it unfortunately neither allows distribution onto separate storage blocks nor high-performance reading with a short-circuit read with its native functions. It can therefore currently only benefit from the advantages of the Hadoop FS for data-parallel processing if the data is saved in contiguous mode and the payload data is accessed directly in raw binary form, bypassing the entire HDF5 software stack for the read operations.

In order to manage the metadata, in contrast to numerical data, the HDF5-based NeXus format (Könnecke, et al., 2015) for neutron, electron and X-ray data sets is currently the most promising solution in the context of LiberTEM. Metadata support and a format definition for pixelated STEM data are under development at the time of writing.

The advantage of NeXus is not so much the encoding of the data, but a strict, well-defined and managed namespace for metadata, including application-specific schema definitions and automated validation tools. This makes the metadata unambiguous and suitable for automated processing across different applications. In principle, the namespace can be encoded in different ways, depending on the application.

## Benchmarking results

LiberTEM was benchmarked on a Supermicro Microcloud 5038MD-H8TRF blade system with the following configuration:

TABLE 1: DESCRIPTION OF THE TEST CLUSTER CONFIGURATION.

| Base system | Supermicro Microcloud 5038MD-H8TRF |
|---|---|
| File system | XFS |
| OS | CentOS 7.6 |
| Kernel | 3.10.0-957.5.1.el7.x86_64 |
| CPU Cluster nodes | Intel(R) Xeon(R) CPU D-1541 @ 2.10GHz |
| CPU Head node | Intel(R) Xeon(R) W-2195 CPU @ 2.30GHz |
| Network interface (cluster) | Intel Ethernet Controller 10G X550T |
| Network interface (Head node) | Intel Ethernet Controller X540-AT2 |
| Switch | ZyXEL XS3700-24 |
| Storage | 2x Samsung SSD 970 EVO 2TB on each node |
| Distributed setup | dask scheduler and dask client in version 1.26.0 running on head node, dask worker running on each cluster node, |
| FIO version | 3.1 |
| Python version | 3.6.6 |
| PyTorch version | 1.0.1.post2 |
| torch backend | mkl |

The raw read performance of the SSD RAID was tested with the "Flexible I/O Tester" (FIO) (Axboe, Flexible I/O Tester) and reached on average 6711 MiB/s, which corresponds precisely to the specified performance of the used SSDs.

IO-bound operation processing performance with LiberTEM was tested on a semi-synthetic 4D STEM dataset for an array of 10,240 x 768 frames of 128 x 128 pixels each, amounting to 480 GiB of float32 values in form of a raw binary file. The CPU-bound operation, i.e. reading from the file system cache, was tested for a semi-synthetic 512 x 768 array of frames with a size of 128 x 128 pixels, amounting to 24 GiB of float32 values. The files were generated from a smaller real 4D STEM data set by tiling it in order to reach a sufficient file size. The files were copied to each node for testing. In production, a suitable file system such as the Hadoop File System, or in the future a caching layer, can be used to use the local storage space more efficiently.

A virtual detector operation with a single random mask was performed on the data set and the time to completion was recorded. The system was using PyTorch with an Intel MKL back-end. The file system caches were dropped for each IO-bound test run and populated for each CPU-bound run.

IO-bound processing throughput with LiberTEM on the cluster scaled nearly linearly with the number of nodes, up to the available maximum number of eight nodes (Figure 8).

The IO-bound processing rate saturated at 5945 MiB/s on a single node, which is approximately 700 MiB/s below the maximum available read rate in FIO. This value was achieved using direct IO which circumvents the file system cache. The use of mmap or conventional file reads achieved lower performance (Figure 9).

CPU-bound processing reached a much higher throughput of 11,407 MiB/s and tapered off only moderately with the number of worker processes until all cores were running a worker process. For comparison, the same CPU-bound test was performed on the high-performance head node with 18 cores (Figure 10).

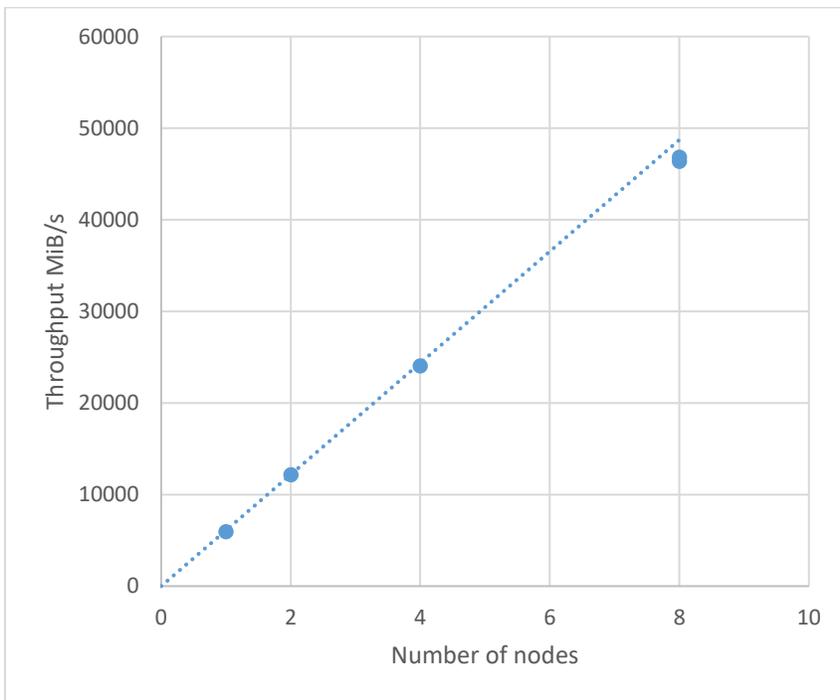

**FIGURE 8: IO-BOUND SCALING BEHAVIOR OF THE TEST CLUSTER DESCRIBED IN TABLE 1 AS A FUNCTION OF THE NUMBER OF PROCESSING NODES. EACH NODE WAS RUNNING WITH EIGHT WORKER PROCESSES, I.E., ONE WORKER PROCESS PER REAL CORE.**

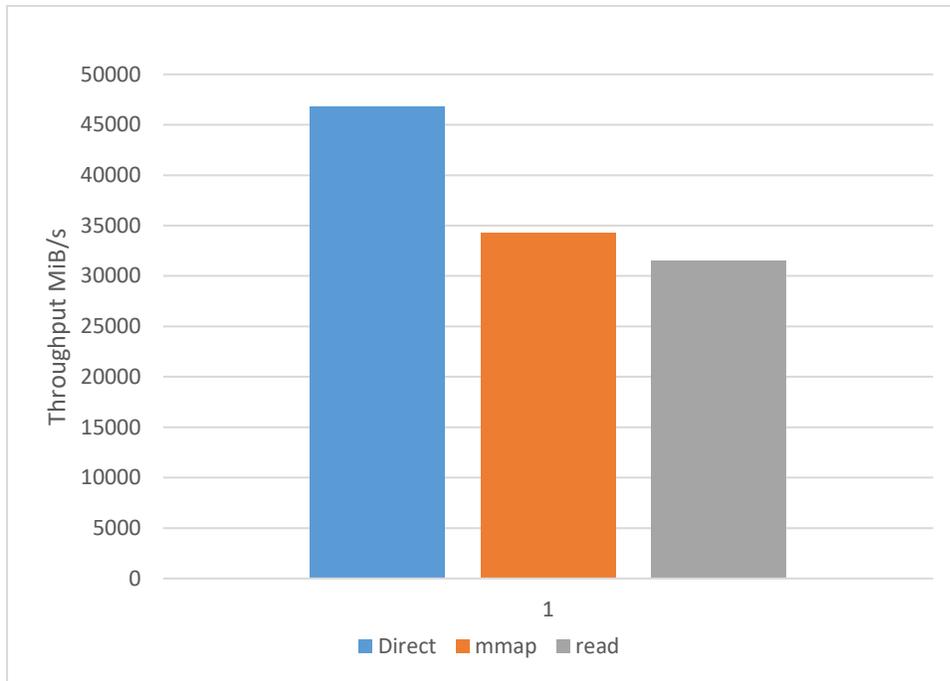

FIGURE 9: COMPARISON OF AGGREGATE PROCESSING RATE OF THE TEST CLUSTER (TABLE 1) WITH READING METHODS TILE-WISE DIRECT IO, MMAP AND CONVENTIONAL TILE-WISE FILE READING IN AN IO-BOUND SCENARIO UNDER MEMORY PRESSURE, I.E. WITH A FILE THAT IS LARGER THAN THE AVAILABLE MEMORY. DIRECT IO AVOIDS THE OVERHEADS ASSOCIATED WITH CONSTANTLY MOVING PAGES IN AND OUT OF THE OPERATING SYSTEM'S PAGE CACHE UNDER SUCH CONDITIONS.

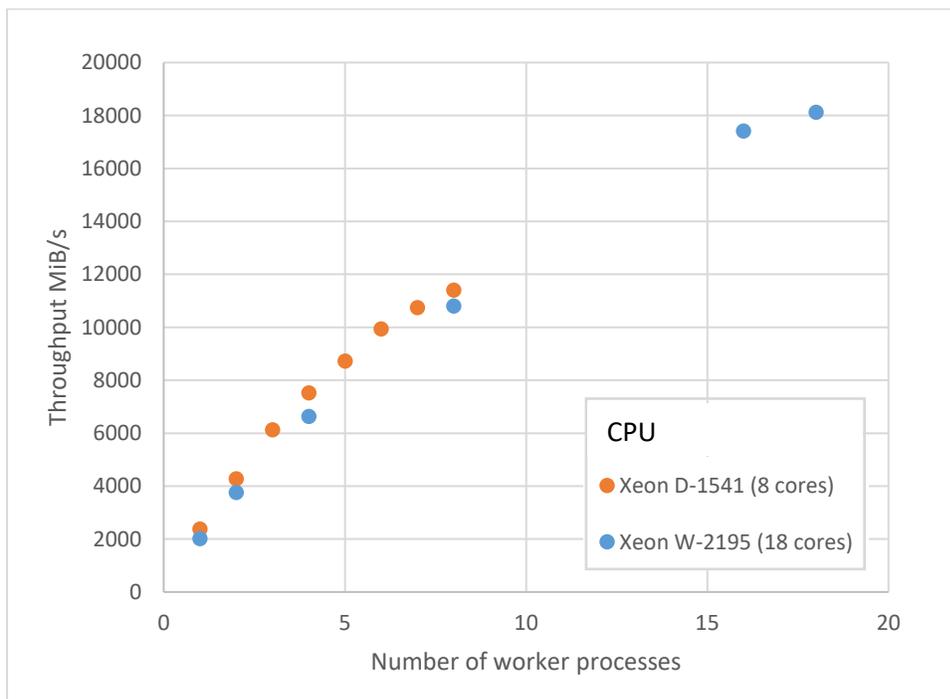

FIGURE 10: COMPARISON OF CPU-BOUND SCALING BEHAVIOR ON A SINGLE PROCESSING NODE AND THE HEAD NODE AS A FUNCTION OF THE NUMBER OF WORKER PROCESSES. MMAP FROM THE FILE SYSTEM CACHE WAS USED FOR READING.

The benchmark results show how an application that is designed for data-parallel processing can benefit from the use of local storage to achieve very high aggregate IO rates on a simple and compact system built from stock components as described in Table 1. A map-reduce-like processing scheme allows very fast and scalable applications to be created with a comparatively simple design.

Achieving such performance required a few targeted optimizations, including eliminating copy operations from the data flow as much as possible. For files that did not fit into memory, direct IO that bypasses the file system cache proved to be most efficient by avoiding overheads from constantly thrashing the cache under heavy memory pressure. Comparison with the FIO benchmarks show that the reader can still be improved, possibly by using a multithreaded reader or advanced interfaces such as liburing (Axboe, Add io_uring IO interface, 2019), which are optimized to read from fast SSDs.

Mmap turned out to be the most efficient method to read from files that can be held in the file system cache. This reading method allows the processing algorithms to operate directly on the file system cache without allocating buffers or making extra copies, if there is enough memory available to hold all of the data.

The linear scaling behavior as a function of the number of processing nodes (Figure 8) shows that the performance can likely be further increased by adding more processing nodes, until the increased load and network traffic on the head node create a bottleneck.

The processing performance on a single node in the CPU-bound scenario scaled well with the number of worker processes up to the number of available physical cores. Hyperthreading provided only a marginal increase. Using a map-reduce-like processing scheme reduces interaction between the workers and makes efficient parallelization easy.

## Development roadmap

For the future, a caching layer that uses local SSDs to cache data from remote sources in an optimized layout without needing the Hadoop file system, a data transformation layer and support for vectored IO are planned.

The processing live data streams during acquisition while streaming them to storage, with the possibility to re-process a growing data set as desired, is in planning as a major update.

Also highly desirable are the application of user-defined functions to a data set. This feature is nearly completed at the time of writing and will likely soon be included in an upcoming release.

Support for physical units and metadata is in the planning stage.

The embedding of LiberTEM in third-party applications is currently at a prototype stage, with a first demonstration achieved for integration in an alpha version of the upcoming Digital Micrograph release by Gatan that supports Python.

## Conclusion and outlook

Building responsive GUI applications for the scalable, distributed processing of very large data sets is a rapidly developing field. It will become a key enabler to work with large data sets that are stored in the cloud, because it allows the processing back-end of such an application to be containerized and executed close to the data storage location, where it can achieve high throughput with little data transfer, while the front-end can even run on a mobile device on a different continent.

The eResearch infrastructure in Australia already implements such workflows, for example coordinated by Microscopy Australia for microscopy applications. Currently, normal desktop applications must be packaged and deployed as virtual machines using a web-based virtual desktop. This layered workaround highlights the gap in requirements between traditional desktop GUI applications that work on local systems and future needs for cloud-based interactive processing. In the future, we expect that more data processing solutions will be developed natively to reap the benefits of distributed storage and processing, while using web technology for their front-ends so that they are ready for cloud deployment.

For electron microscopy, responsive and interactive processing of live detector data with well-defined open programming interfaces, convenient user interfaces, high throughput and very short response times below 100 ms will be a key towards future applications where automation and algorithms are working hand in hand with human interaction and data interpretation. Dynamic instrument control may even require response loops with only a few milliseconds turn-around.

Detector data with a high information content can be analyzed by using forward models in a feedback loop that compare acquired data from multiple detectors with simulations based on models of the specimen. This approach is already used with very good results to reconstruct the magnetization states of samples (Caron, 2018) (Song, et al., 2018). The implementation of such an analysis would require a hybrid system that combines fast iterative optimization of a model with the handling of a massive data set. If it can be scaled up to provide real-time operation, then it can inform a microscope what data to acquire next, in order to gather required information to refine a sample model to a desired level of precision.

In the future, we envision an integrated solution that combines the use of correlative methods, simulation, visualization, instrument control and the full automation of workflows in a suitable system architecture. Such a system can be built of interoperable components from different scientific and commercial contributors. Open standards and open interfaces, in combination with Open Source core components, can enable the development of such an advanced solution.

A heterogeneous system of many components connected using a glue language leaves potential for optimization. For performance-critical parts, a framework that handles distributed computation and data locality (such as Spark) and allows easy development and interfacing (like Python) but is compiled down to machine code and incorporates full-scale integrated optimization, from high level variable elimination, blocking and data locality to cache-efficient processing and optimal machine instructions, would be desirable.

Improvements in the infrastructure for developing complex high-performance distributed applications are currently tasks for computer scientists and IT specialists. We hope that in the future scientists will again be able to implement their analysis workflows *ad hoc* using simple tools, without a deep knowledge of the underlying mechanisms. Systems such as LLVM Polly (Grosser, et al.), Apache Spark, Dask.distributed and LiberTEM are already working in this direction.

## Funding

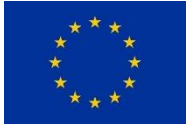
This project has received funding from the European Research Council (ERC) under the European Union's Horizon 2020 research and innovation programme (grant agreement No 780487).

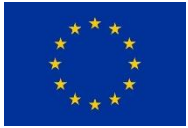
This project has received funding from the European Union's Horizon 2020 research and innovation programme under grant agreement No 823717 – ESTEEM3.

The Ernst Ruska Centre at the Jülich Research Center supports development through employing key contributors and providing the required infrastructure for development and testing.

## Acknowledgments

The authors would like to acknowledge the ESTEEM3 network, many colleagues at Forschungszentrum Jülich and other institutions, and developers and users of various software and hardware solutions for valuable insights and discussions. https://libertem.github.io/LiberTEM/acknowledgments.html